\documentclass[11pt]{article} 
\pdfoutput=1 

\usepackage{jheppub} 

\usepackage[utf8]{inputenc}
\usepackage[english]{babel}
\usepackage{amsmath,amssymb,amsbsy,amstext,amsthm,booktabs,simplewick,exscale,relsize,slashed,graphicx,amsfonts,upgreek, xcolor,subcaption}
\usepackage{multirow,color,dcolumn,bm,enumerate}
\usepackage{hyperref}
\usepackage{feynmp-auto,axodraw2,tikz}
\usepackage{ulem}
\usepackage{comment}

\usepackage{hyperref}
\usepackage[capitalise,noabbrev,nameinlink]{cleveref}

\title{Non-factorizable Superamplitudes for Massive $\mathcal N=1$ Superstates}
\author[a]{Antonio Delgado,}
\author[a]{Adam Martin,}
\author[a]{Runqing Wang}

\affiliation[a]{Department of Physics and Astronomy, University of Notre Dame,
  South Bend, IN, 46556 USA}

\emailAdd{adelgad2@nd.edu}
\emailAdd{amarti41@nd.edu}
\emailAdd{rwang7@nd.edu}

\abstract{In this paper we study non-factorizable $\mathcal{N}=1$ superamplitudes for massive chiral superstates. We demonstrate how little group scaling and  the supersymmetric Ward identities determine the form of non-factorizable massless superamplitudes, then extrapolate to massive superamplitudes by requiring they reduce to the massless form when we send all masses to zero. This technique does not depend on whether or not the superstates are self-conjugate (so that the fermionic components are either Dirac or Majorana) or whether the superamplitude is dressed with a form-factor. }

\begin{document}
\maketitle
\setcounter{page}{2}

\section{Introduction}
\label{intro}
Spinor-helicity formalism allows for a purely on-shell formulation of computing S-matrix elements in four dimensions. Even though a complete framework where one can solely work with the on-shell formulation is far from reality, the construction for all masses and spins in non-supersymmetric theories is well known~\cite{Shadmi:2018xan,Ma:2019gtx,Aoude:2019tzn,Durieux:2020gip,DeAngelis:2022qco,AccettulliHuber:2021uoa,Liu:2023jbq} and is now a standard approach to calculating scattering amplitudes for the LHC. By enlarging the spacetime symmetry to include fermionic generators, supersymmetry can be naturally encoded in the formalism which allows one to calculate  superamplitudes. Many results have been derived in $\mathcal{N}=4$ Super Yang-Mills (SYM) theory, e.g. maximally helicity violating (MHV) and next-to-MHV (NMHV) superamplitudes \cite{Elvang:2013cua}, dual superconformal symmetry \cite{Drummond:2008vq}, supersymmetric Ward identity and its general solutions \cite{Elvang:2009wd, Elvang:2010xn}, etc. 

The off-shell formulation of $\mathcal{N}=1$ supersymmetry in four dimensions has been studied for decades \cite{Wess:1992cp}, and it is the only possibility  if  Nature chooses to be supersymmetric at a high energy scale. The on-shell construction of $\mathcal{N}=1$ SYM has received  some attention~\cite{Bern:1994zx,Bern:1994cg,Bern:1997ng, Bern:2002tk, DeFreitas:2004kmi, Bidder:2005ri,Britto:2006sj, Bern:2009xq}, usually being regarded as a reduction from $\mathcal{N}=4$ case \cite{Elvang:2011fx,Herderschee:2019dmc}. Moving beyond SYM theories, all possible three-particle superamplitudes with massive chiral and vector superfields were constructed in \cite{Herderschee:2019ofc}, with an emphasis on renormalizable interactions.

 If one wants to work with effective field theory, non-renormalizable interactions are indispensable. Therefore, the goal of this paper is to study how to calculate superamplitudes in $\mathcal N=1$ supersymmetry from higher dimensional operators, i.e. the non-factorizable $N$-point pieces, where $N$ is the number of particles involved in the amplitude. We will focus on non-factorizable amplitudes involving chiral and anti-chiral supermultiplets. 

In previous works~\cite{Delgado:2024ivu, Delgado:2023ogc}, we constructed non-factorizable superamplitudes of massless chiral states using a map, or `replacement rule' between higher dimensional operators in the Lagrangian and (supersymmetric) spinor-helicity variables. This was inspired by a similar treatment in non-supersymmetric theories, such as in the Standard Model Effective Field Theory~\cite{Shadmi:2018xan,Ma:2019gtx,Henning:2019mcv,Henning:2019enq,Durieux:2019siw,Liu:2023jbq}. However, that method is not particularly intuitive for superamplitudes of massive states, as supersymmetric Lagrangians are conventionally written in using pairs of chiral superfields (short representations of the supersymmetry algebra) connected by a mass term in the potential, rather than working in terms of the full massive multiplets (long representations).

Therefore, in this paper we begin by revisiting the construction of non-factorizable superamplitudes of massless states using dimensional analysis, little group scaling, and supersymmetric Ward idenitites (Sect~\ref{sec:massless}). We identify the scenarios where the amplitude is unique (Sect.~\ref{massless_nonfac}) versus when there are multiple amplitudes that fit the criteria (Sect.~\ref{sec:ffactornomass}). We call the former `base amplitudes', which -- in the language of higher dimensional operators -- correspond to higher dimensional operators with no derivatives. Adding derivatives -- which becomes a momentum prefactor to the base amplitude -- introduces more flexibility and admits more possible amplitudes, which we can interpret as the amplitude version of partitioning derivatives in an operator multiple ways. Once we understand how mass dimension and little group scaling constrain massless superamplitudes, we can extrapolate to the massive case (Sect.~\ref{massivecase}) by requiring that superamplitudes of massive states reproduce the massless case in the appropriate limit.  We work through several examples, illustrating how our technique matches results in the literature, discuss how the self-conjugate nature of the superstate affects their superamplitudes, then end with a discussion of future directions (Sect.~\ref{sec:conclusions}). Same technical details can be found in the appendices.

\section{Massless superamplitudes:}\label{sec:massless}
\subsection{Basics of massless (super) spinor-helicity} \label{sec:masslessintro}

The basic ingredients in superamplitudes are superstates, which are built from coherent states via the introduction of a Grassmann parameter $\eta_i$, $|\eta_i \rangle = |s+1/2 \rangle + \eta_i |s \rangle$, where $s$ is the spin of the state (and we have suppressed the state's momentum dependence).  Taking an inner product with $\langle p|$  (and $s=0$, such that the highest spin is 1/2) we can write $\Phi = \psi + \eta\, \phi$, where  $\Phi = \langle p|\eta \rangle$ is the coherent state wave function (and $\phi = \langle p |s=0 \rangle, \psi = \langle p |s = 1/2\rangle$). 

 To connect $\phi$, $\psi$ to off-shell superfields, which we'll indicate in bold throughout, we identify them with the contraction of a field on a state, so $\phi = \langle p |  \bm{ \hat \phi} |0 \rangle \sim 1$, $\psi = \langle p |  \bm{\hat \psi}_\alpha |0 \rangle \sim \lambda_\alpha$, where $\lambda_\alpha$ is the usual positive helicity spinor helicity variable.~\footnote{Here $\alpha, \dot\alpha$ are Lorentz indices, and we have stripped off the exponential factor which leads to overall momentum conservation.}  Lumped together, we see we can identify the states forming $\Phi$ -- a superstate built from the highest spin $1/2$ -- with  the components of an on-shell chiral superfield.
 
 We can also build a coherent state with highest spin zero, $|0\rangle + \eta |-1/2 \rangle$, which we identify with an on-shell anti-chiral superfield; $\Phi^\dag = \phi^* + \eta \psi^\dag$,  with $\psi^\dag = \langle p |\bm{\hat \psi^{\dag,\dot{\alpha}}}|0\rangle \sim \tilde \lambda^{\dot{\alpha}}$, the negative helicity spinor-helicity variable.   Inspecting these, we see the `wave functions'  or `on-shell components' have mass dimension $[\phi] =[\phi^*]=  0$, $[\psi] = [\psi^\dag] = 1/2$.\footnote{The reader may notice $\eta$ appears differently in the $\Phi$ vs. $\Phi^\dag$ wave function. As the supercharges $\mathcal Q^\dag$  and $\mathcal Q$ do not commute, we cannot have our states be eigenstates of both. We have chosen coherent states to be eigenstates of $\mathcal Q^\dag$. This choice, called the `$\eta$' basis, leads to the differences in $\Phi$ vs. $\Phi^\dag$.  The `$\eta^\dag$' basis, where coherent states are eigenstates of $\mathcal Q$, is equally valid. In most circumstances, we can derive results in either basis, though in certain circumstances explained in Ref.~\cite{Delgado:2023ogc}, only one choice gives sensible results. }
 
 The supercharges $\mathcal Q, \mathcal Q^\dag$ can be expressed in terms of spinor-helicity variables and $\eta$,
 \begin{align}
 \mathcal Q_\alpha = \sum\limits_{i=1}^N \lambda_{i,\alpha} \frac{\partial}{\partial \eta_i}, \quad\quad \mathcal Q^{\dag,\dot{\alpha}} = \sum\limits_{i=1}^N \tilde\lambda^{\dot{\alpha}}_i \eta_i
 \end{align}
 where $i$ labels a particular superstate in an amplitude and we sum over all states when forming the charges. We will sometimes refer to $\mathcal Q = \sum\limits_i \mathcal Q_i, \mathcal Q^\dag  = \sum\limits_i \mathcal Q^\dag_i$, where $\mathcal Q_i, \mathcal Q^\dag_i$ are the supercharge projections along the momentum of particle $i$.

A major advantage of working with spinor-helicity variables is that they explicitly reveal an amplitude’s little group properties. Little group transformations leave the momentum of a given state $i$ unchanged, which is accomplished in spinor helicity variables by $\lambda_i \to t_i \lambda,_i \tilde \lambda_i \to t^{-1}_i \tilde \lambda_i$, for a unit complex number $t_i$. Extending this scaling to $\eta_i \to t_i \eta_i$ leaves the supercharges (both the sum, and the individual $\mathcal Q_i, \mathcal Q^\dag_i$)  unaffected. It also makes the little group scaling of the on-shell superfields covariant, as $\eta_i \to t_i \eta_i$ and $\psi_i \sim \lambda_i \to t_i\, \lambda_i$ mean every $\Phi_i$ in the amplitude scales as $t_i$ (and each $\Phi^\dag$ is unaffected). Explicitly,
\begin{align}
\mathcal A(\Phi_i, \cdots \Phi_{N_c}, \Phi^\dag_{N_c+1}, \cdots \Phi^\dag_{N_c+\bar N_c}) \to t_1\, t_2,\cdots t_{N_c}\mathcal A(\Phi_i, \cdots \Phi_{N_c}, \Phi^\dag_{N_c+1}, \cdots \Phi^\dag_{N_c+\bar N_c}).\nonumber
\end{align}

One drawback of the spinor-helicity construction is that it only knows about kinematics and little group scaling. Global symmetries, either continuous or discrete that are manifest from a field perspective aren't automatic and must be imposed by hand (meaning we would use it to allow or forbid a given amplitude). For supersymmetric theories, R-symmetry  also falls into this category. We set the R-charges such that $R[\theta]=-R[\bar{\theta}]=1$ and $R[\Phi]=R_\Phi$. For example, in the massless Wess-Zumino model with a single superfield, $R_\Phi=\frac{2}{3}$.  The R-charge of a superstate is defined to be the same as its highest component, and we define $R[\eta]=1$ to allow all components in a superstate carry the same R-charges.

\subsection{Non-factorizable amplitudes from higher dimensional operators} \label{massless_nonfac}

Consider a higher-dimensional D-term operator of the form $\mathbf{ (D\bar D)^{2N_D}\,\Phi^{N_c}\, (\Phi^\dag)^{\bar N_c}}$ (with all fields assumed to be distinguishable bosonic superfields for now) and $N_c, \bar N_c \ge 2$. In the Lagrangian, this term is suppressed by $\Lambda^{N_c + \bar N_c + 2N_D -2}$. This operator will make a non-factorizable contribution to an amplitude with $N_c + \bar N_c $ external legs, an object that carries mass dimension $4-(N_c + \bar N_c)$.

Within the $\eta$ basis, this contribution can be written as 
\begin{align}
\mathcal A(\Phi_i, \cdots \Phi_{N_c}, \Phi^\dag_{N_c+1}, \cdots \Phi^\dag_{N_c+\bar N_c}) = \frac{1}{\Lambda^{N_c + \bar N_c + 2N_D -2}}\, \delta^{(2)}(\mathcal Q^\dag)\, F(\lambda_i, \tilde \lambda_i, \eta_i) \nonumber
\end{align}
where $\lambda, \tilde \lambda, \eta$ are spinor-helicity variables for the $N_c + \bar N_c$ legs and $\delta^{(2)}(\mathcal Q^\dag)$ is the Grassmann delta function
\begin{align}
\delta^{(2)}(\mathcal Q^\dag) =  \sum\limits_{i<j}^N \langle i j \rangle \eta_i \eta_j.
\end{align}

 Using the fact that $\delta^{(2)}(\mathcal Q^\dag) \sim \langle ij \rangle$ and that the $\eta_i$ are dimensionless, we can equate the mass dimension of the expression above to $4-(N_c + \bar N_c)$ to solve for $[F]$, the mass dimension of $F(\lambda, \tilde\lambda, \eta)$. We find
\begin{align}
[F] = 1 + 2N_D. \nonumber
\end{align}
Note that the dimension of $F$ is independent of the number of chiral/anti-chiral superfields.

The form of $F$ is further constrained by little group scaling and the supersymmetric Ward identity $\mathcal Q \mathcal A = 0$. As $\delta^{(2)}(\mathcal Q^\dag)$ is little group invariant, the little group scaling must be carried by $F$. Similarly, as $ Q \delta^{(2)}(\mathcal Q^\dag)= 0$, the Ward identity requires $\mathcal Q F = 0$. 

Let's set $N_D = 0$ for now and begin with $N_c = 2$, so an operator of the form $\bm{\Phi_1\Phi_2 \Phi^{\dag}_3 \cdots \Phi^{\dag}_{\bar N_c + 2}}$. We can expand $F$ in powers of $\eta_i$, where each term carries the little group scaling $t_1 t_2$
\begin{align}
F = a_1\, \eta_1 \eta_2 + b^\alpha_1 \eta_1 \lambda_{2,\alpha} + b^\alpha_2\, \eta_2 \lambda_{1,\alpha} + c_1 [12] + c^{\alpha\beta}_2 \lambda_{1,\alpha} \lambda_{2,\beta}. 
\label{eq:Fexpand}
\end{align}
We emphasize that the restriction to non-factorizable amplitudes plays a key role here, as it means that the little group scaling must be set in the numerator, e.g. we cannot achieve the correct scaling for $\Phi_1 \Phi_2$ via $1/\langle 12 \rangle$. In Eq.~\eqref{eq:Fexpand}, the coefficients $a_1 - c_2$ must be little group neutral, so they are either constant or functions of $\eta_i \tilde \lambda_i$ or $\lambda_i \tilde \lambda_i$. From this, we see that $b_1 = b_2 = c_2 = 0$ as there is no way to make a Lorentz invariant without pushing $[F]$ past 1. From dimensional arguments alone, $c_1 = const$ is still possible, as is $a_1 = a_{ij} \eta_i \eta_j \langle ij \rangle, i,j = 3 \cdots \bar N_c + 2$  (as $\eta^2_1 = \eta^2_2 = 0$) and $i \ne j$.~\footnote{The  possibility, $a_1 = \lambda_i \tilde \lambda_i, i = 1\cdots N_c + \bar N_c$, is not Lorentz invariant. Adding extra spinors to make it Lorentz invariant pushes the mass dimension too high.}

Now we impose the Ward identity $\mathcal Q F = 0$. For $c_1 = const$ this is trivially satisfied, while for $a_1$, we find
\begin{align}
\label{eq:WardI}
0 = (\eta_2\, \lambda_1 - \eta_1  \lambda_2)\, a_{ij} \eta_i \eta_j \langle ij \rangle + \eta_1\,\eta_2\, a_{ij}  \langle ij \rangle (\eta_j \lambda_i - \eta_i \lambda_j).
\end{align}
To satisfy this, each $\eta$ combination must separately vanish, something that is only possible if $a_{ij} = 0 \to a_1 = 0$.  Therefore, purely using on dimensional grounds and little group scaling, we know
\begin{align}
\mathcal A(\Phi_1, \Phi_2, \Phi^\dag_{3}, \cdots \Phi^\dag_{N_c+\bar N_c})_{N_D = 0} = \frac{c}{\Lambda^{ \bar N_c }}\, \delta^{(2)}(\mathcal Q^\dag)\, [12]\nonumber
\end{align}
where $c$ is a dimensionless constant.

Now let's extrapolate to $N_c$ chiral superfields, an operator of the form $\bm{\Phi_1\cdots \Phi_{N_c}\Phi^\dag_{\bar N_c+1}\cdots \Phi^{\dag}_{\bar N_c + N_c}}$. Expanding out $F$ in powers of $\lambda_i$, there will be terms with zero, one, and two $\lambda_i$:
\begin{align}
\label{ex:expandF3}
F =& a_1 \eta_1 \cdots \eta_{N_c} + (b^\alpha_1 \eta_2 \cdots \eta_{N_c} \lambda_{1,\alpha} + perms) + \nonumber \\
& (c_{12}\, \eta_3 \cdots \eta_{N_c} [12] + perms) + (d^{\alpha\beta}_{12}\, \eta_3 \cdots \eta_{N_c}\, \lambda_{1,\alpha}\lambda_{2,\beta} + perms)
\end{align}
The terms with one $\lambda$ or with two uncontracted $\lambda$ can't be made Lorentz invariant without exceeding $[F]=1$, while the terms with zero $\eta$ will lead to conditions similar to Eq.~\eqref{eq:WardI} upon applying the Ward identity which can only be satisfied if $a_1 = 0$. This leaves $c_{ij} = const$ as the only possibility. 

There are $\binom{N_c}{2}$ constants $c_{ij}$. Applying the Ward identity to these terms, we get $\binom{N_c}{2}(N_c-2)$ terms grouped into $\binom{N_c}{N_c-3}$ distinct polynomials containing $(N_c-3)$ $\eta$ powers. Each polynomial multiplies a sum of 
$\binom{N_c}{2}(N_c-2)/{\binom{N_c}{N_c-3}} = 3\, c_{ij}$ factors and is subject to 2 conditions from the Ward identity, leaving only one coefficient. As each $c_{ij}$ appears in $(N_c-2)$ different $\eta$ polynomials, there are enough conditions to completely determine the amplitude up to an overall coefficient.
\begin{align}
\mathcal A(\Phi_1, \cdots, \Phi_{N_c}, \Phi^\dag_{N_c+1}, \cdots, \Phi^\dag_{N_c+\bar N_c})_{N_D = 0} = \frac{c}{\Lambda^{ N_c +\bar N_c -2}}\, \delta^{(2)}(\mathcal Q^\dag)\, (\eta_3 \cdots \eta_{N_c} [12] + perms).\nonumber
\end{align}
Tallying up the total $\eta$ scaling, we see that $\mathcal A$ is a degree $N_c$ polynomial in $\eta$ (which we'll write as $\eta^{N_c}$). 

Any $F$ of the form $\mathcal Q^2(G)$ for some $G(\lambda, \tilde\lambda, \eta)$\footnote{Here, the Lorentz indices of the $\mathcal Q$ are contracted, so $\mathcal Q^2 = \mathcal Q^\alpha \mathcal Q_\alpha$. We will often neglect the spinor indices.} will automatically satisfy $\mathcal Q F = 0$. As the form of $F$ is unique, at least for $N_D=0$, and since $F  = \mathcal Q^2(G)$ is guaranteed to work, it must be the case that $F  = \mathcal Q^2(G)$ here -- we just need to find the $G$ that matches the forms we find.  $\mathcal Q^2$ carries no little group weight but has mass dimension 1; therefore, to recover the correct little group scaling for $F$, $G$ must be a product of each $\eta_i$ only. Indeed, one can verify that
\begin{align}
\label{eq:Qform}
F = \mathcal Q^2(\eta_1\eta_2 \cdots \eta_{N_c})
\end{align}
matches the forms shown in the examples above.\footnote{Using this form, we can justify the restriction $N_c, \bar N_c > 2$. The amplitude has the form 
\begin{align}
\delta^{(2)}(\mathcal Q^\dag) \mathcal Q^2(\eta^{N_c}) =  \mathcal Q^2(\delta^{(2)}(\mathcal Q^\dag)\eta^{N_c}) 
\end{align}
given that $\mathcal Q$ annihilates the $\delta$-function. Recall $\delta^{(2)}(\mathcal Q^\dag)$ is $\mathcal O(\eta^2)$, with $\eta$ indices running from $1\cdots N$, the total number of particles in the amplitude. If $N < N_c + 2$, all of the terms in the $\delta$-function include one $\eta_i$ with index $i \in 1\cdots N_c$, so the net $(\delta^{(2)}(\mathcal Q^\dag)\eta^{N_c}) \propto \eta^2_i =0$. If $N \ge N_c + 2$, implying $\bar N_c \ge 2$, there are enough different $\eta_i$ that $(\delta^{(2)}(\mathcal Q^\dag)\eta^{N_c})$ is non-vanishing. We emphasize that this restriction only applies to non-factorizable terms.}

This form for the $N_D = 0$ amplitude is reproduced with the following rules (assuming $N_c > 2$):
\begin{enumerate}
\item[1.)] For each $\Phi_i$ leg in the amplitude, include a factor of $\eta_i$. This carries the appropriate little group scaling and (on-shell) mass dimension.
\item[2.)] For each $\Phi^\dag_i$ leg include a factor of $1$. Again, this carries the appropriate little group scaling and (on-shell) mass dimension.
\item[3.)] Apply $\mathcal Q^2$ to the product of $\eta_i$.
\item[4.)] Multiply by $\delta^{(2)}(\mathcal Q^\dag)$.
\end{enumerate}
In Ref.~\cite{Delgado:2023ogc} we put forth a similar looking mapping, except that it connected off-shell, Lagrangian superfields and on-shell amplitudes:
\begin{align}
\bm{\Phi_i} \to \eta_i,\quad \bm{\Phi^\dag_i} \to 1,\quad \int d^4\theta \to \delta^{(2)}(\mathcal Q^\dag) \mathcal Q^2.
\end{align}
This Lagrangian $\to$ amplitude mapping was inspired by work in non-supersymmetric theories~\cite{Ma:2019gtx, Henning:2019mcv, Henning:2019enq}. For massless superfields, either mapping suffices, however for amplitudes involving massive superparticles the former form is more transparent. The transparency comes from the fact that, even when fields are massive, we work with a Lagrangian written in terms short multiplets (chiral and anti-chiral superfields) with a mass in the potential, rather than working with long multiplets.\footnote{This is not just a supersymmetry issue, we do the same thing in SMEFT -- e.g. working with chiral fields and connecting them via mass terms in the potential}

\subsection{Form factors in massless superamplitudes}\label{sec:ffactornomass}
 Now let us add superderivatives. What changes? The dimension of $F$ increases by $2N_D$, but the little group scaling of any given field is unchanged. To see the latter, we can apply (on-shell) $\mathcal Q_i$ to (on-shell) $\Phi_i$,
\begin{align}
\mathcal Q_i\Phi_i = \lambda_i \frac{\partial}{\partial \eta_i}(\eta_i + \lambda_i) = \lambda_i \nonumber
\end{align}
which scales $(\mathcal Q_i\Phi_i) \to t_i (\mathcal Q_i\Phi_i)$ just as $\Phi_i$ did. The same holds for $\mathcal Q^\dag_i$ applied to $\Phi^\dag_i$. Additionally, the combination $\mathcal Q^\dag \mathcal Q$ doesn't change the {\it overall} (meaning independent of $i$ index) $\eta$ scaling -- $\mathcal Q$ removes an $\eta$, while $\mathcal Q^\dag$ puts it back. This does not mean that the index $i$ remains the same after $\mathcal Q^\dag \mathcal Q$, just the net exponent.

 Something similar happens in non-supersymmetric field theories.  An operator $\bm{ \partial^m\phi^n}$  contributes to the same amplitude -- in terms of number of legs and little group scaling -- as $\bm{\phi^n}$, just with a momentum dependent (and therefore little group invariant) form factor.~\footnote{In the language of Ref.~\cite{Liu:2023jbq}, $\phi^n$ generates a Stripped Contact Term, which is dressed by additional Mandelstam invariants from the $\partial^m$ factors.} To compensate for the form factor, the mass dimension of the amplitude must change. 

The form factor analogy will serve us well when we turn to massive superamplitudes. There, it is less useful to think in terms of off-shell operators (with or without derivatives), and instead think about a minimum amplitude, meaning the smallest $[F]$, plus a form factor which we can expand in a tower of increasing $[F]$. In a non-supersymmetric theory, the only little group invariant we have around is the momentum $\lambda_{i} \tilde \lambda_i$. In supersymmetry, we have two others: $\lambda_{i} \partial/\partial\eta_{i}$ and $\tilde \lambda_{i}\eta_i$.

Given that there are multiple little group invariants, and the fact that one of them acts derivatively, one may expect that enumerating the possibilities (at a given mass dimension) is more complicated than in a non-supersymmetric theory. However, as shown in Appendix~\ref{app:Qproof}, non-factorizable amplitudes including a form factor continue to satisfy
\begin{align}
\label{eq:generalform}
F  \propto  \mathcal Q^2( \eta^{N_\eta}),
\end{align}
where $\eta^{N_\eta}$ is shorthand for a polynormial of degree $N_\eta$. Here $N_\eta$ is some integer, which does not a priori have to be equal to the number of chiral superfields (compare with Eq.~\eqref{eq:Qform}).

In the absence of a form factor,  the mass dimension of $F$ is saturated by the $\mathcal Q^2$ so the entire little group scaling must be carried by the $\eta$ factors. In this case, it is clear that $N_\eta = N_c$, and the proportionality factor between the two sides of Eq.~\eqref{eq:generalform} must be a (dimensionless) constant. 

After including a form factor, such that the mass dimension of $F$ can be greater than 1, we can carry the little group weight of a chiral superfield with an extra $\lambda_i$ outside the $\mathcal Q$ . To keep the $\eta$ power consistent, the `missing' $\eta$ must come from (meaning carry the index of) one of the non-chiral superfields. Each $\eta$ we replace this way must be accompanied by a $\tilde \lambda$ to keep the little group scaling consistent. Said another way, $N_\eta = N_c$ in this case as well, but which $\eta_i$ appear may be different than when there is no form factor.

As an example, consider an amplitude with three chiral superfields and two anti-chiral fields, $\mathcal A(\Phi_1, \Phi_2, \Phi_3, \Phi^\dag_4, \Phi^\dag_5)$. The only form consistent with $[F] = 1$ is
\begin{align}
\mathcal A \propto  \delta^{(2)}(\mathcal Q^\dag) \mathcal Q^2(\eta_1\eta_2\eta_3). \nonumber
\end{align}
If we consider $[F]$ with mass dimension three, 
\begin{align}
\mathcal A \propto  \delta^{(2)}(\mathcal Q^\dag)\,s_{ij}\, \mathcal Q^2(\eta_1\eta_2\eta_3), i,j \in 1\cdots 5, i\ne j \nonumber
\end{align}
are allowed. However, so are terms with $\eta_4, \eta_5$ within the $\mathcal Q^2$ argument, such as 
\begin{align}
\mathcal A \propto  \delta^{(2)}(\mathcal Q^\dag)\langle i 4\rangle [3i]\mathcal Q^2(\eta_1\eta_2\eta_4), i \in 1 \cdots 5 \nonumber 
\end{align}
or
\begin{align}
\mathcal A \propto  \delta^{(2)}(\mathcal Q^\dag)\langle 4 5\rangle [32]\mathcal Q^2(\eta_1\eta_4\eta_5). \nonumber
\end{align}

In Ref~\cite{Delgado:2023ogc, Delgado:2024ivu}, we showed that amplitudes with form factors can be derived by extending the replacement rule mentioned earlier to include derivatives acting on fields (or, staying on-shell, derivatives on external amplitude legs). Specifically, we can view the different form factors as different partitioning of derivatives among the amplitude legs. For a given partitioning:
\begin{enumerate}
\item Follow the original rules for each $\Phi_i$ and $\Phi^\dag_i$ leg.
\item For each $\mathcal Q \Phi_i$ leg, include a factor of $\lambda_j (\partial/\partial \eta_j)\, \eta_i = \lambda_i$. For more derivatives, e.g. $\mathcal Q^\dag Q \Phi_i$, continue to apply the on-shell expressions for $\mathcal Q, \mathcal Q^\dag$; $\mathcal Q^\dag \mathcal Q \Phi_i \to \eta_i \tilde \lambda_i \lambda_i$, etc.
\item Following the same logic for $\mathcal Q^\dag \Phi^\dag_i$, $\mathcal Q^\dag \Phi^\dag_i \to \eta_i \tilde \lambda_i$, $\mathcal Q \mathcal Q^\dag \Phi^\dag_i \to \tilde \lambda_i \lambda_i$.
\item Apply $\mathcal Q^2$ to the product of $\eta_i$ and add the $\delta^{(2)}(\mathcal Q^\dag)$.
\end{enumerate}

Given the faithfulness of these rules, it is useful to list the $\mathcal Q$ or $\mathcal Q^\dag$ on the appropriate leg. Taking a few of the examples from above,
\begin{align}
\label{eq:qamp}
\mathcal A(\mathcal Q^\dag\mathcal Q \Phi_1, \mathcal Q^\dag\mathcal Q \Phi_2, \Phi_3, \Phi^\dag_4, \Phi^\dag_5) &\propto \delta^{(2)}(\mathcal Q^\dag)\,s_{12}\, \mathcal Q^2(\eta_1\eta_2\eta_3) \nonumber \\
\mathcal A(\mathcal Q^\dag\mathcal Q \Phi_1,  \Phi_2, \mathcal Q \Phi_3, \mathcal Q^\dag \Phi^\dag_4, \Phi^\dag_5) & \propto \delta^{(2)}(\mathcal Q^\dag)\langle 1 4\rangle [31]\mathcal Q^2(\eta_1\eta_2\eta_4) \\
\mathcal A( \Phi_1,  \mathcal Q\Phi_2, \mathcal Q \Phi_3, \mathcal Q^\dag \Phi^\dag_4, \mathcal Q^\dag \Phi^\dag_5) & \propto  \delta^{(2)}(\mathcal Q^\dag)\langle 4 5\rangle [32]\mathcal Q^2(\eta_1\eta_4\eta_5). \nonumber
\end{align}
Note that the $\mathcal Q, \mathcal Q^\dag$ in the amplitude label are just a way to differentiate between different form factors dressing the same `base' amplitude, $\mathcal A(\Phi_1, \Phi_2, \Phi_3, \Phi^\dag_4, \Phi^\dag_5)$. Further, the amplitudes derived this way above are not all independent. As shown in Ref.~\cite{Delgado:2024ivu}, a complete basis for the amplitudes corresponding to any given operator class can be determined by exploiting a hidden $U(N)$ symmetry and using Young Tableau techniques.

\section{Massive superamplitudes:} \label{massivecase}
\subsection{Basics of massive (super) spinor-helicity}

We can apply many of the lessons learned studying massless amplitudes directly to the massive superfield case.  

The major change in moving from massless to massive amplitudes is that the little group for massive particles is $SU(2)$, such that amplitudes involving spin $|s|$ fields are symmetric $2s$ index tensors under the representative $SU(2)$ group. For non-supersymmetric theories, the spinor-helicity variables $\lambda_I, \tilde \lambda_I, I= 1,2$ are the only objects that carry little group indices (and therefore the amplitudes must be tensors of these alone), while in the supersymmetry case the $\eta$ are also promoted to $SU(2)$ doublets and can be used to form symmetric products.  Appendix~\ref{app:massive review} contains a brief review of massive spinors and some useful identities involving massive spinors and $\eta_I$. For a more thorough introduction to massive spinor helicity constructions, see Ref.~\cite{Arkani-Hamed:2017jhn, Durieux:2019eor}.

The promotion of $\eta$ to a $SU(2)$ doublet also makes sense from the coherent state picture, as massive $\mathcal N=1$ supersymmetry has $2$ creation and annihilation operators. As such, acting on an initial state (a Clifford vacuum) with spin $s = 0$, the multiplet contains a second $s=0$ state and state of spin $1/2$. In terms of wave functions, and working in the $\eta$ representation (where superstates are eigenstates of $\mathcal Q^\dag$) we can write:
\begin{align}
\label{eq:massstate}
\Psi = -\frac 1 2 \eta_I \eta^I \phi + \eta_I \chi^I + \tilde \phi,
\end{align}
where we note all pieces are $SU(2)$ singlets, the little group scaling for $s=0$. The individual component wave functions have the same dimensionality as in the massless case, $[\phi] = [\tilde \phi] = 0$, $[\chi^I] = 1/2$. 

The supercharges can be written in terms of the (supersymmetric) spinor helicity variables 
\begin{align}
 \mathcal Q_\alpha = \sum\limits_{i=1}^N \lambda_{iI,\alpha}\frac{\partial}{\partial \eta_{iI}},\quad\quad \mathcal Q^\dag_{\dot\alpha} = \sum\limits_{i=1}^N \tilde \lambda_{iI, \dot\alpha}\, \eta^{I}_i, \nonumber
\end{align}
with $N$ equal to the number of superstates in the amplitude. As in the massless case, all superamplitudes must be annihilated by both supercharges.\footnote{For massive $\mathcal N = 1$ supersymmetry, the Grassmann delta function becomes $\delta^{(2)}(\mathcal Q^\dag) = \Big( \sum\limits_{i<j}^N \langle i^I j^J\rangle \eta_{iI}\eta_{iJ} + \frac 1 2 \sum\limits_{i=1}^N m_i \eta_{iI}\eta^I_i \Big)$.}

As written, Eq.~\eqref{eq:massstate} assumes there is no relation between $\phi$ and $\tilde \phi$ (or between $\lambda_+$ and $\lambda_-$). If we take these to be complex conjugates of each other, $\Psi$ represents a massive Majorana state and $\bar{\Psi} = \Psi$. Otherwise, the state is Dirac, meaning that the conjugate state $\bar{\Psi}$ is distinct from $\Psi$ so we need a second superstate to exhaust all degrees of freedom.  In terms of component wave functions, we need two complex scalars $\phi_a, \phi_b$ and two Weyl fermions $\chi_a, \chi_b$ grouped as:
\begin{align}
\label{eq:dirac}
\Psi =  -\frac 1 2 \eta_I \eta^I \phi_a + \eta_I \left\{ \begin{array}{c}\chi^I_a \\ \chi^{\dag I}_b \end{array} \right\} + \phi^*_b,\quad\quad \bar\Psi =  -\frac 1 2 \eta_I \eta^I \phi_b + \eta_I \left\{ \begin{array}{c}\chi^I_b \\ \chi^{\dag I}_a \end{array} \right\}+  \phi^*_a  
\end{align}
We emphasize here that although we write superstates as $\Psi$ and $\bar\Psi$, it doesn't mean that the two are hermitian conjugate.

In the massless limit, a self-conjugate (Majorana fermion component) $\Psi$ falls apart into a chiral plus an anti-chiral state.  
\begin{align}
\Psi|_{i,m=0} = \hat\eta\,\Phi_i  + \Phi^\dag_i \nonumber
\end{align}
which we get by replacing $\eta_+ \to \hat\eta, \eta_- \to \eta$ in Eq.~\eqref{eq:massstate}\footnote{To see this explicitly, we take $\eta_I = (\eta_+, \eta_-)$ and $\chi_{iI} = \{\lambda_{iI} , \tilde \lambda_{iI}\}$. See Appendix~\ref{app:massive review} for more detail on massive spinor helicity variable and their reduction to the massless case}.
In terms of on-shell components, this becomes
\begin{align}
\Psi_{i,m=0}  = \hat\eta_i(\eta_i + \lambda_i) + (1 + \eta_i\tilde\lambda_i). \nonumber
\end{align}
We can pick off either the chiral or anti-chiral piece by taking $\partial/\partial\hat\eta_i$ or by setting $\hat\eta_i = 0$.

For a non self-conjugate $\Psi$, the story is similar, except there are two distinct chiral superfields in the massless limit:
\begin{align}
\Psi|_{m=0} = \hat\eta\,\Phi_{a}  + \Phi^\dag_b, \quad\quad \bar\Psi|_{m=0} = \hat\eta\,\Phi_{b}  + \Phi^\dag_a,
\end{align}
where $\Phi_{a,b}$ have component fields as in Eq.~\eqref{eq:dirac}.

In Sec.~\ref{massless_nonfac}, we introduced a replacement rule for massless amplitudes as a shortcut to the form argued by general principles. This form involved replacing each chiral /anti-chiral leg in the amplitude with its lowest dimension component, $\Phi_i \to \eta_i$, $\Phi^\dag_i \to 1$. Taking the lowest dimension component of a massive superstate, then taking the massless limit, 
\begin{align}
\Psi_{i,lowest} = 1 - \frac 1 2 \eta_{iI}\eta^{I}_i \xrightarrow[m=0]{}  \hat \eta_i  \eta_i+ 1
\end{align}
we (taking $\partial/\partial\hat\eta_i$ or $\hat\eta_i = 0$) recover the massless superstate replacement rules. Note that for a Dirac state, we need an additional label to keep track of which of the two massless fields ($(a,b) $  in the notation above) we mean, as the $\hat \eta$ component of $\Psi$ gives $\Phi_a$, distinct from the $\hat \eta$ component of $\Psi^\dag$.

Finally, massive superstates can have an R-symmetry, though there is less freedom than in the massless. For a massive Majorana state, the mass term sets $[R_\Psi] = 1$, while for a Dirac state the mass allows arbitrary R-charge provided  $[R_\Psi] + [R_{\bar{\Psi}}] = 2$. While R-symmetry played no role in our discussion of massless superamplitudes\footnote{Specifically, for a single Lagrangian term we can always choose R-charges to allow the interaction}, we will see in the next section that it can do more for us in the massive case.

\subsection{Massive Superamplitudes}\label{sect:massivenoff}
We have reviewed the basics of massive super spinor-helicity formalism and we are prepared to construct the superamplitudes. We will first consider self-conjugate superstates (Majorana fermions), then determine what must be adjusted for the non self-conjugate case. 

From our experience with massless superfields, we expect that $F = \mathcal Q^2(G)$ where the minimum (meaning no form factor) $[F] =1$. $G$ must i.) have mass dimension zero, ii.) carry the appropriate little group scaling for each particle involved ($SU(2)$ singlet for amplitudes formed from Eq.~\eqref{eq:massstate} objects), iii.) reproduce the massless little group scaling for any given particle if we take the high energy limit (where, in the massless limit,  $\Psi$ will fall apart into both/either massless $\Phi, \Phi^\dag$).  The unique form that satisfies these requirements is 
\begin{align}
 G = \prod\limits_{i=1}^N (1 - \frac{1}{2}\eta_{iI}\eta^I_i). \nonumber 
\end{align}
Taking only the constant $1$ piece only captures the anti-chiral piece of $\Psi$ in the massless limit, while the $-\frac 1 2\eta_{iI}\eta^I_i$ term only captures the chiral piece. To ensure that the 'daughter' amplitudes containing chiral or anti-chiral pieces of $\Psi|_{m=0}$ are related to each other and to the `mother' amplitude before taking the massless limit, both $SU(2)$ singlets must appear and have the same coefficient, hence the $(1 - \frac{1}{2}\eta_{iI}\eta^I_i)$ factors.

Adding the factors of $\mathcal Q$ to convert $G$ to $F$ and including the overall $\mathcal Q^\dag$ delta function, the amplitude is
\begin{align}
\label{eq:massiveA}
\mathcal A(\Psi_1 \cdots \Psi_N) = \delta^{(2)}(\mathcal Q^\dag)\mathcal Q^2\Big( \prod\limits_{i=1}^N(1 - \frac{1}{2}\eta_{iI}\eta^I_i)\Big). 
\end{align}
 As an example, take $N=3$, so $\mathcal A(\Psi_1, \Psi_2, \Psi_3)$: 
\begin{align}
\label{eq:AAA}
\mathcal Q^2 \left( \prod_{i=1}^3(1-\frac{1}{2}\eta_{iI}\eta^I_i)\right)&=
(\sum\limits_{i=1}^3 \lambda_{iI}\frac{\partial}{\partial \eta_{jI}})^2[(1-\frac{1}{2}\eta_{1J}\eta^J_1)(1-\frac{1}{2}\eta_{2K}\eta^K_2)(1-\frac{1}{2}\eta_{3M}\eta^M_3)] \nonumber \\
&\sim \{\frac{1}{2}m_1(\eta_2^2+\eta_3^2-2)+[1^I2^J]\eta_{1I}\eta_{2J}\}+perm.
\end{align}
where we have introduced the shorthand $\eta^2_i =\eta_{iI}\eta^I_i$\footnote{We've also suppressed spinor indices on the $\mathcal Q$ for brevity. In actual calculations these should be reinstated}. In the last line, we've dropped the term $\propto \eta^2_2 \eta^2_3$ as it vanishes when hit by $\delta^{(2)}(\mathcal Q^\dag)$. This reproduces the 3-pt amplitude $\mathcal A'(\Psi_1,\Psi_2,\Psi_3)$ in \cite{Herderschee:2019ofc} once we use the conservation of $\mathcal Q^\dag$, i.e. $\tilde{\lambda}_{1I}\eta^I_1+\tilde{\lambda}_{2I}\eta^I_2+\tilde{\lambda}_{3I}\eta^I_3=0$ to eliminate $\eta^I_{3}$  (see Appendix.~\ref{remove eta} for details):
\begin{equation}
\label{eq:threeptv2}
\mathcal A(\Psi_1,\Psi_2,\Psi_3)\propto\delta^{(2)}(\mathcal Q^\dag)\left[-m_3+\left([1^I2^J]\eta_{1I}\eta_{2J}+\frac{1}{2}(m_2\eta_{1I}\eta_1^I+m_1\eta_{2I}\eta_2^I)\right)\right].
\end{equation}
Comparing our result with the result in Ref. \cite{Herderschee:2019ofc}: 
\begin{equation}
\mathcal A'(\Psi_1,\Psi_2,\Psi_3)\propto\delta^{(2)}(\mathcal Q^\dag)\left[\lambda+b\left([1^I2^J]\eta_{1I}\eta_{2J}+\frac{1}{2}(m_2\eta_{1I}\eta_1^I+m_1\eta_{2I}\eta_2^I)\right)\right],
\end{equation} 
we see that in our case $\lambda=-bm_3$ or $b=-\frac{\lambda}{m}$ if all three masses are the same $m_1=m_2=m_3=m$ and this also matches the result in \cite{Herderschee:2019ofc} if we ignore the possible non-trivial intrinsic parity phases.\footnote{In principle, the two parameters $\lambda$ and $b$ are not related because each piece individually satisfies the little group scaling and Ward identity. However, if we go to the massless limit and identify one piece with the Hermitian conjugate of the other piece, then the two parameters are no longer independent.} 

The three-particle superamplitude above matches the result from the Wess-Zumino model with purely renormalizable interactions, despite the fact that our construction is designed for (in off-shell language) higher dimensional operators. Our result matches the renomalizable case because the three-particle amplitudes constructed from scalars and fermions alone have no denominator and can therefore be thought of as 'non-factorizable'. For superamplitudes with more than three particles, there will be generally be factorizable contributions whose form is not captured by Eq~\eqref{eq:massiveA}. 

So far, we have assumed that the states $\Psi$ are Majorana, so $\bar{\Psi} = \Psi$. For Dirac, $\Psi \ne \bar{\Psi}$, so we need two superstates and one may expect the formalism to change. However, for the purposes of constructing massive superamplitudes, no change is needed -- the construction as describes holds for both $\Psi$ and $\bar{\Psi}$, and we only need to keep track of $\Psi$ vs. $\bar{\Psi}$ for the purposes of projecting out components or taking the massless limit. As an example, consider the four-particle, non-factorizable amplitude involving two Majorana states (labeled with subscript $M$), one Dirac $\Psi$ and one $\bar\Psi$:
\begin{align}
A_4(\Psi_{M1},\Psi_{M2},\Psi_{3},\bar\Psi_4) &\sim \delta^2(Q^\dag)\mathcal Q^2\Big(\prod_{j=1}^4(1-\frac{1}{2}\eta_j^2)\Big) \nonumber \\
&\sim\delta^2(Q^\dag)\{m_1(-\frac{1}{2}\eta_2^2+1)(-\frac{1}{2}\eta_3^2+1)(-\frac{1}{2}\eta_4^2+1)\\
&-[1^I2^J]\eta_{1I}\eta_{2J}(-\frac{1}{2}\eta_3^2+1)(-\frac{1}{2}\eta_4^2+1)+perm(1234)\} \nonumber
\label{eq:4pt}
\end{align}
This is just Eq.~\eqref{eq:massiveA} with $N=4$. The Dirac nature of the states becomes apparent from the diversity of amplitudes in the $m_3, m_4 \to 0$ limit:
\begin{align}
A_4(\Psi_{M1},\Psi_{M2},\Phi_{3a},\Phi_{4b})  &\sim \delta^2(Q^\dag)\mathcal Q^2\Big(\prod_{j=1}^2(1-\frac{1}{2}\eta_j^2)\,\eta_3\,\eta_4\Big) \nonumber \\
A_4(\Psi_{M1},\Psi_{M2},\Phi^\dag_{3b},\Phi_{4b})  &\sim \delta^2(Q^\dag)\mathcal Q^2\Big(\prod_{j=1}^2(1-\frac{1}{2}\eta_j^2)\,\eta_4\Big)  \\
A_4(\Psi_{M1},\Psi_{M2},\Phi_{3a},\Phi^\dag_{4a})  &\sim \delta^2(Q^\dag)\mathcal Q^2\Big(\prod_{j=1}^2(1-\frac{1}{2}\eta_j^2)\,\eta_3\Big) \nonumber \\
A_4(\Psi_{M1},\Psi_{M2},\Phi^\dag_{3b},\Phi^\dag_{4a})  &\sim \delta^2(Q^\dag)\mathcal Q^2\Big(\prod_{j=1}^2(1-\frac{1}{2}\eta_j^2)\Big) \nonumber 
\end{align}
Note that amplitudes containing $\Phi_{3a},\Phi_{4a}$ or $\Phi_{3b},\Phi_{4b}$ do not occur here, but would show up had we considered $A_4(\Psi_{M1},\Psi_{M2},\Psi_{3},\Psi_{4})$.

One of the features we noted in our review of massless amplitudes was the overall scaling with $\eta$ -- meaning the total number of $\eta$ appearing. In the massless case, we saw that amplitudes scaled as $\eta^{N_c}$. Inspecting Eq.~\eqref{eq:massiveA} for the analogous concept for massive amplitudes, we see that an amplitude with $N$ massive superstates scales as $A_2 \eta^2  + A_4 \eta^4 + \cdots A_{2N-2} \eta^{2N-2}$, where $\eta^2$, etc. stand for some $SU(2)$ singlet contraction on $\eta_I$ and $A_i$ are $\eta$-independent factors. This general counting is verified in Eq.~\eqref{eq:threeptv2}, where $N=3$ and we see terms $\mathcal O(\eta^2)$ and $\mathcal O(\eta^4)$.

Unlike massless case, R-symmetry can help reduce this general $A_2 \eta^2  + A_4 \eta^4 + \cdots A_{2N-2} \eta^{2N-2}$ form. The R-charge is fixed by the mass type, so the first question is whether we impose that symmetry on all other interactions. In Majorana case, $R=1$, so -- if we enforce R-symmetry -- all amplitudes should scale as $\eta^{N_\Psi}$. For any odd $N_\Psi$, this isn't possible as the general form for $\mathcal A$ contains only even $\eta$ powers. This is just the amplitude-level statement of the fact that $[R]=1$ is inconsistent with either superpotential or Kahler interactions with an odd number of fields. For the Dirac case, R-symmetry requires $[R_\Psi] + [R_{\bar{\Psi}}] = 2$, so there is more flexibility to allow interactions. As an example, consider a single species of massive chiral states, taking $[R_\Psi] = 0, [R_{\bar{\Psi}}] = 2$.  If we impose $R$ symmetry on the three-particle interaction,  then we must have that 
\begin{align}
\mathcal A(\Psi, \Psi, \Psi) \propto \eta^0 & \quad\quad \mathcal A(\Psi, \Psi, \bar\Psi) \propto \eta^2 \nonumber \\
\mathcal A(\Psi, \bar\Psi, \bar\Psi) \propto \eta^4 & \quad\quad \mathcal A(\bar\Psi, \bar\Psi, \bar\Psi) \propto \eta^6 \nonumber 
\end{align}
The $\mathcal A(\Psi, \Psi, \Psi) $ and $\mathcal A(\bar\Psi, \bar\Psi, \bar\Psi)$ scalings are not possible given the $\eta$ power counting consistent with supersymmetry so these amplitudes must be zero, while $\mathcal A(\Psi, \Psi, \bar\Psi) $ and $\mathcal A(\Psi, \bar\Psi, \bar\Psi)$ pick out pieces of Eq.~\eqref{eq:threeptv2}.

\subsection{Form factors for massive superamplitudes}\label{sec:ffactormassive}

Now let's add a form factor, assuming our states to be Majorana for simplicity. We are guided by three criteria. First, as in the massless case (Eq.~\eqref{eq:qamp}), we can dissect the form factor by viewing it as applying $\mathcal Q, \mathcal Q^\dag$ to the legs. Each $\mathcal Q, \mathcal Q^\dag$ should add mass dimension $+1/2$ to the amplitude. Furthermore, the form factor should not change the little group scaling of the state, so it should be an $SU(2)$ singlet. Finally, we need $\mathcal Q \Psi$ and $\mathcal Q^\dag \Psi$ to reproduce the massless results, namely
\begin{align}
\mathcal Q\Psi_i \to \left\{ \begin{array}{cc} \lambda_i & \text{for}\, \Phi_i \\ 0 & \text{for}\, \Phi_i^\dag \end{array} \right. \quad \mathcal Q^\dag \Psi_i \to \left\{ \begin{array}{cc} 0 & \text{for}\, \Phi_i \\ \tilde \lambda_i \eta_i & \text{for}\, \Phi_i^\dag \end{array} \right. 
\end{align}
The only forms consistent with all three criteria are: $\mathcal Q\Psi_i \to \lambda_{iI}\eta^I_i, \mathcal Q^\dag \Psi_i \to \tilde \lambda_{iI} \eta^{I}_i$. Of course, the net form factor must be some combination of pairs of $\mathcal Q\Psi_i$,  $\mathcal Q^\dag\Psi_i$ as a single instance of either is not Lorentz Invariant.

 Using this technique, we can generate the massive analogs of Eq.~\eqref{eq:qamp}:
\begin{align}
\mathcal A(\mathcal Q^\dag\mathcal Q \Psi_1, \mathcal Q^\dag\mathcal Q \Psi_2, \Psi_3, \Psi_4, \Psi_5)  &\propto \delta^{(2)}(\mathcal Q^\dag) [1_I 2_J]\langle 1_K 2_L \rangle  \mathcal Q^2\Big(\eta^I_1 \eta^J_2 \eta^K_1 \eta^L_2\,\prod\limits_{i=1}^5(1 - \frac{1}{2}\eta_{iM}\eta^M_i)\Big) \nonumber \\
\mathcal A(\mathcal Q^\dag\mathcal Q \Psi_1, \Psi_2, \mathcal Q \Psi_3, \mathcal Q^\dag \Psi_4, \Psi_5)  &\propto \delta^{(2)}(\mathcal Q^\dag) [1_I 3_J]\langle 1_K 4_L \rangle  \mathcal Q^2\Big(\eta^I_1 \eta^J_3 \eta^K_1 \eta^L_4\,\prod\limits_{i=1}^5(1 - \frac{1}{2}\eta_{iM}\eta^M_i)\Big)  \nonumber \\
\mathcal A( \Psi_1, \mathcal Q\Psi_2, \mathcal Q \Psi_3, \mathcal Q^\dag \Psi_4, \mathcal Q^\dag\Psi_5)  &\propto \delta^{(2)}(\mathcal Q^\dag) [2_I 3_J]\langle 4_K 5_L \rangle  \mathcal Q^2\Big(\eta^I_2 \eta^J_3 \eta^K_4 \eta^L_5\,\prod\limits_{i=1}^5(1 - \frac{1}{2}\eta_{iM}\eta^M_i)\Big)  \nonumber 
\end{align}

\section{Conclusions}\label{sec:conclusions}

Amplitudes provide with an alternative way to formulate a QFT which is purely on-shell and therefore free of the redundancies that appear in the standard Feynman approach to perturbation theory. 

In this paper we have given the most general form of the non-factorizable piece of a $N$-point massive superamplitude with (anti) chiral superstates. We constructed these amplitudes using dimensional analysis, little group scaling, and by requiring that they correctly reproduce the massless superamplitude results in the appropriate limit. The little group scaling and supersymmetric Ward identities are not sufficient to completely pin down the superamplitude, as we can always attach a momentum dependent `form factor' , the amplitude version of the fact that $\phi^n$ and  $\partial^m \phi^n$ both contribute to to an $n$-scalar process.  However, just as one can explore different derivative partition in an operator like $\partial^m \phi^n$, we can form the superamplitude for any given form factor.  The master formulae can be found in in  Sects.~\ref{sect:massivenoff} (no form factor) and ~\ref{sec:ffactormassive} (including a form factor).

A possible application of this technique is to provide an alternative way to construct a complete basis for non-renormalizable operators in a supersymmetric effective field theory of massive superfields. The redundancies coming from the equation of motion are absent when one uses the (super)amplitude formalism since everything is on-shell and the equations of motion are trivial.

An obvious follow up is to include vectors superpermultiplets into the picture which is the topic of a forthcoming publication.  A potential complication for vector superfields is that massless three-point amplitudes involving vectors and originating from renormalizable interactions have powers of momentum in the denominator, while our construction assumes momentum factors can only sit in the numerator.

\acknowledgments

We thank Yael Shadmi and Seth Koren for helpful comments.
This is partially supported by the National Science Foundation under Grant Numbers PHY-2112540 and PHY-2412701.

\appendix 

\section{Proof that all non-factorizable amplitudes contain a factor of $\mathcal Q^2(\eta_i\cdots)$}
\label{app:Qproof}

Forgetting any assumptions about derivatives, dimensions, or little group scaling, we can expand a non-factorizable amplitude in powers of $\eta$, using a basis set $b_i(\eta^k)$ to span polynomials of degree $k$ in $\eta$. For example, let $k=3, N=4$ and the basis is given by $b_1=\eta_1\eta_2\eta_3,b_2=\eta_2\eta_3\eta_4,b_3=\eta_3\eta_4\eta_1,
b_4=\eta_4\eta_1\eta_2$.
\begin{align}
\label{eq:amp1}
\mathcal A = \delta^{(2)}(\mathcal Q^\dag) \sum_{i,k} a_{ik}(\lambda, \tilde \lambda)\, b_i(\eta^k)
\end{align}
Here the sum $k$ runs from $0$ to $N_\eta$, some number less than or equal to $N$ (not yet necessarily equal to the number of chiral superfields, the number of particles involved in the amplitude, while $i$ runs from $1$ to $\binom{N_\eta}{k}$. To satisfy the Ward identity, we need
\begin{align}
\mathcal Q\mathcal A = \sum_{i,k} a_{ik}(\lambda, \tilde \lambda)\sum\limits_{j=1}^{N} \lambda_j b_{ij}(\eta^{k-1}) = 0, \nonumber
\end{align}
where $k$ runs from $1$ to $N_\eta -1$, $i$ from $i$ to $\binom{N_\eta}{k-1}$ and $j$ runs $1$ to $N$\footnote{While $\mathcal Q = \sum\limits_{i=1}^N \lambda_i \partial/\partial \eta_i$, the $\eta$ in $b$ only run from $1\cdots N_\eta$, so in effect the upper limit of the sum in $\mathcal Q\,b$ is also $N_\eta$}. This must vanish for each $i$ and $k$, as different polynomials in $\eta$ are independent, so
\begin{align}
\sum_j a_{ik}(\lambda, \tilde \lambda) \lambda_j = 0. \nonumber 
\end{align}
For generic $\lambda, \tilde \lambda$, this sum can only vanish if it is proportional to the total momentum $\sum_j \lambda_j \tilde \lambda_j$, or if each of the $a_{ik}(\lambda, \tilde \lambda) \lambda_j$ are proportional to $\chi_{ijk} = [ij]\lambda_k + [ki]\lambda_j + [jk]\lambda_i$, which vanishes for any $i\ne j \ne k$ by the Schouten identity. Proportionality to the total momentum is not actually viable, as that would imply $a_{ik}(\lambda, \tilde\lambda)$ contains an un-contracted factor of $\tilde \lambda_j$, but such a factor would mean $a_{ik}$, and thus $\mathcal A$, is not Lorentz invariant.  That leaves $\chi_{ijk}$ as the only choice. \begin{align}
\mathcal Q \mathcal A = \sum_{ijk}\chi_{ijk\,}b_{ij}(\eta^{k-1}) = 0.
\end{align}
In order for $a_{ik}(\lambda, \tilde \lambda) \lambda_j$ to form a  Schouten product, the indices of all the $\lambda_i$ must all have the same range; $\lambda_j$ is restricted to $1 \cdots N_\eta$ so the $\lambda, \tilde \lambda$ in $a$ must also lie in this range.

To have enough Schouten products so that each $b_{ij}$ has a vanishing prefactor, we need
\begin{align}
\binom{N_\eta}{k-1} = \binom{N_\eta}{3}.
\end{align}
The right hand side counts the groups of three $\lambda_i$, $i \in 1\cdots N_\eta$ (rather than $N$) for reasons emphasized earlier, while the left hand side is the number of polynomials involving $k-1$  powers of $\eta_i, i \in 1 \cdots N_\eta$. The two sides are equal if $k = N_\eta -2$.\footnote{Another solution is $k=4$, which we dismiss as it is independent of the number of superstates.} Thus, the  $k$ in Eq.~\eqref{eq:amp1} collapses and we only need to consider $\eta$ polynomials of length $N_\eta-2$. 

Next, note that $\chi_{ijk} = \mathcal Q^3(\eta_i\eta_j\eta_k)$ (making it immediately obvious why it vanishes), so
\begin{align}
\label{eq:q3}
\mathcal Q \mathcal A = \sum_{ij} \mathcal Q^3(\eta_i\eta_j\eta_k)b_{ij}(\eta^{N_\eta-3}).
\end{align}
The $\eta$ indices within the argument of $\mathcal Q^3()$ and for the basis polynomials in $b_{ij}$ all lie in $1\cdots N_\eta$. Furthermore, these indices must be distinct, as $\eta^2_i = 0$. Thus, we can lump the $b_{ij}$ into the argument of $\mathcal Q^3$: $\mathcal Q^3(\eta^3)b_{ij}(\eta^{N_\eta-3}) \sim \mathcal Q^3(\eta^{N_\eta})$. Stripping off one $\mathcal Q$ from Eq.~\eqref{eq:q3} to get the amplitude, we have
\begin{align}
\mathcal A \propto  \delta^{(2)}(\mathcal Q^\dag) \sum \mathcal Q^2(\eta^{N_c}).
\end{align}
As $\mathcal Q$ removes one power of $\eta$, each term in the sum as $N_\eta$ powers of $\eta$, as expected. These combine with the $\eta$ factors in the delta function to give a net scaling of $\mathcal A \sim \eta^{N_\eta}$.

Let us check that this is indeed the case for the $N_\eta=4$ case. Let $m_{ijk}\equiv Q^2(\eta_{ijk})$ and we have the following four terms:
\begin{equation}
A_1=m_{123}\eta_4,\ \ A_2=m_{234}\eta_1,\ \ A_3=m_{341}\eta_2,\ \ A_4=m_{412}\eta_3,
\end{equation}
and they are related by: $A_1-A_2+A_3-A_4=Q^2(\eta_{1234})\equiv Q^2(\eta^4)$. Note that if we choose a different linear combination, then the result cannot be annihilated by Q. We conclude that $QA=0$ uniquely fixes the form of the superamplitude up to an overall $\lambda(\tilde\lambda)$-dependent factor when $N_\eta=4$. 

\section{Massive spinor-helicity formalism}\label{app:massive review}
We show the conventions and notations we use in this paper. For a detailed introduction to massless spinor-helicity formalism, one may refer to \cite{Elvang:2013cua}, and \cite{Arkani-Hamed:2017jhn, Durieux:2019eor} for massive case. For the most part, we follow Ref.~\cite{Durieux:2019eor}, though we swap the definition of the angle and square brackets, using $\lambda = |i]$ for positive helicity (in the massless limit) and $\tilde \lambda = |i \rangle$ for negative helicity. We use a convention where the Lorentz indices for negative helicity spinors $\tilde\lambda$ and little group $SU(2)$ indices are contacted lower-to-upper, $\,_{\dot{\alpha}} \,^{\dot{\alpha}}$ or $\,_I \,^I$, while Lorentz indices for positive helicity spinors $\lambda$ are contracted upper-to-lower $\,^{\alpha} \,_{\alpha}$.

A massive momentum can be decomposed into 
\begin{equation}
p^{\dot{\alpha}\beta}=|p^{I\dot{\alpha}}\rangle[p_I^\beta|,
\end{equation}
where $I$ is the index of massive $SU(2)$ group. Some useful properties of these spinor momenta are given by:
\begin{align}
&\langle p^Ip^J\rangle = m\epsilon^{IJ},\ [p^Ip^J]=-m\epsilon^{IJ}\\
&|p_I]_\alpha[p^I|^\beta=m\delta^\beta_\alpha,\ |p_I]_\alpha\langle p^I|_{\dot{\beta}}=p_{\alpha\dot{\beta}},\ |p_I\rangle^{\dot{\alpha}}\langle p^I|_{\dot{\beta}}=-m\delta_{\dot{\beta}}^{\dot{\alpha}},\ |p_I\rangle^{\dot{\alpha}}[p^I|^\beta=-p^{\dot{\alpha}\beta}
\end{align}
Two identities one immediately gets from the above are:
\begin{align}
&(\lambda_{I}\frac{\partial}{\partial\eta_I})(\lambda_{J}\frac{\partial}{\partial\eta_J})(-\frac{1}{2}\eta^2+1)=\lambda^I\lambda_I=2m\\
&[i_Ij_J](\lambda_{kI}\frac{\partial}{\partial\eta_{kI}})\eta_i^I\eta_j^J=m_i\lambda_j^J\eta_{jJ}+m_j\lambda_i^I\eta_{iI}
\end{align}

To make the massless limit of massive simple, it is convenient to choose the spatial direction of the (massive) momentum along the direction of the corresponding particle's motion. Following Ref.~\cite{Arkani-Hamed:2017jhn}, we can decompose
\begin{align}
p_\mu = k_\mu + q_\mu, 
\end{align}
with
\begin{align}
k_\mu = \frac{E+P}{2}(1,0,0,1), \quad q_\mu = \frac{E-P}{2}(1,0,0,-1).
\end{align}
Here, $q$ scales as $m^2/E$ and therefore vanishes in the massless limit. With this momentum decomposition, we can identify
\begin{align}
\lambda_{iI} = \left( \begin{array}{c} -|k_i ] \\ |q_i ] \end{array} \right), \quad\quad \tilde{\lambda}_{iI} = \left( \begin{array}{c} |q_i \rangle \\ |k_i \rangle \end{array} \right),
\end{align}
so that, in the massless limit
\begin{align}
\lambda_{iI} \to \left( \begin{array}{c} - \lambda_i \\ 0 \end{array} \right), \quad\quad \tilde{\lambda}_{iI} \to \left( \begin{array}{c} 0 \\ \tilde \lambda_i \end{array} \right).
\end{align}

\section{Removing $\eta_3$}\label{remove eta}
We now show that \eqref{eq:AAA} reduces to \eqref{eq:threeptv2}. Let us first invoke the super Ward identity,
\begin{equation}
Q^\dag=\tilde{\lambda}_{1I}\eta_1^I+\tilde{\lambda}_{2I}\eta_{2}^I+\tilde{\lambda}_{3I}\eta_{3}^I=0,
\end{equation}
we can solve for $\eta_3$ in terms of other variables:
\begin{equation}\label{eta3}
\eta_3^I=-\frac{1}{m_3}(\langle3^I1^J\rangle\eta_{1J}+\langle3^I2^J\rangle\eta_{2J})
\end{equation}
Let's calculate each term one at a time by substituing \eqref{eta3}:
\begin{equation}
\begin{split}
&[2_I3_J]\eta_2^I\eta_3^J\\
=&-\frac{1}{m_3}[2_I3_J]\eta_2^I\left(\langle3^J1^K\rangle\eta_{1K}+\langle3^J2^K\rangle\eta_{2K}\right)\\
=&-\frac{1}{m_3}\left([2_I3_J]\langle3^J2^K\rangle\eta_2^I\eta_{2K}+[2_I3_J]\langle3^J1^K\rangle\eta_2^I\eta_{1K}\right)\\
=&-\frac{1}{m_3}\left(-[2_I3_J]\langle3^J2_K\rangle\frac{1}{2}\epsilon^{IK}\eta_{2L}\eta_{2}^L-[2_I1_J]\langle1^J1^K\rangle\eta_2^I\eta_{1K}-[2_I2_J]\langle2^J1^K\rangle\eta_2^I\eta_{1K}\right)\\
=&\frac{1}{m_3}\left(p_2\cdot p_3\eta_2^2+m_1[2_I1_J]\eta_2^I\eta_1^J-m_2\langle 2^I1^J\rangle\eta_{2I}\eta_{1J}\right)
\end{split}
\end{equation}
Similarly,
\begin{equation}
[3_I1_J]\eta_3^I\eta_1^J=\frac{1}{m_3}\left(p_1\cdot p_3\eta_1^2+m_2[1_I2_J]\eta_1^I\eta_2^J-m_1\langle 1^I2^J\rangle\eta_{1I}\eta_{2J}\right),
\end{equation}
and the one with masses:
\begin{equation}
\begin{split}
&\frac{1}{2}(m_1+m_2)\eta_3^2\\
=&\frac{m_1+m_2}{2m_3^2}\left(\langle 3_J1^I\rangle \eta_{1I}+\langle 3_J2^I\rangle \eta_{2I}\right)\left(\langle 3^J1^K\rangle \eta_{1K}+\langle 3^J2^K\rangle \eta_{2K}\right)\\
=&\frac{m_1+m_2}{2m_3^2}(m_1m_3\eta_1^2+m_2m_3\eta_2^2+2\langle3_J1^I\rangle\langle3^J2^K\rangle\eta_{1I}\eta_{2K}).
\end{split}
\end{equation}

Put everything together we get:
\begin{equation}\label{without eta3}
\begin{split}
&A_3(\Phi_1,\Phi_2,\Phi_3)\\\propto&\{\frac{1}{2}m_1(\eta_2^2+\eta_3^2-2)+[1_I2_J]\eta_1^I\eta^J_2\}+perm\\
=&-(m_1+m_2+m_3)+[1_I2_J]\eta^I_1\eta^J_2+\frac{1}{2}(m_1+m_3)\eta_2^2+\frac{1}{2}(m_2+m_3)\eta_1^2\\
&+\frac{1}{2}(m_1+m_2)\eta_3^2+[2_I3_J]\eta^I_2\eta^J_3+[3_I1_J]\eta^I_3\eta^J_1\\
=&-(m_1+m_2+m_3)+\frac{m_1m_2+m_1m_3+m_2^2+m_3^2+2p_2 \cdot p_3}{2m_3}\eta_2^2\\
&+\frac{m_1m_2+m_2m_3+m_1^2+m_3^2+2p_1 \cdot p_3}{2m_3}\eta_1^2+\frac{m_1+m_2+m_3}{m_3}[1_I2_J]\eta^I_1\eta^J_2\\
=&\frac{m_1+m_2+m_3}{m_3}\{-m_3+[1_I2_J]\eta^I_1\eta^J_2+\frac{1}{2}(m_1\eta_2^2+m_2\eta_1^2)\},
\end{split}
\end{equation}
where we use the conservation of momentum $\sum_i p_i=0$ on the last line.

\section{Massless limits of $\mathcal A(\Psi_1,\Psi_2,\Psi_3)$}\label{app:mless}
As an additional check on our formulation, we can verify that the high energy limit of Eq.~\eqref{eq:AAA} recovers what we expect. As our form reproduces the results of Ref.~\cite{Herderschee:2019ofc}, and that reference explored the massless limits, we expect no surprises. Nevertheless, the limits serve to illustrate how to manipulate our more 'symmetric' construction of the amplitude (meaning we don't use the $\mathcal Q^\dag$ constraint to remove any $\eta_i$). 

Taking $m_1 \to 0$ and retaining the chiral part of $\Psi_1$ by sending $1-\frac{1}{2}\eta_{1K}\eta^K_1 \to \eta_1$:
\begin{align}
\label{eq:m1tozerochiral}
\mathcal A(\Phi_1, \Psi_2, \Psi_3) &\propto \delta^{(2)}(\mathcal Q^\dag) (\sum\limits_{i=1}^3 \lambda_{iI}\frac{\partial}{\partial \eta_{jI}})^2[\eta_1\,(1-\frac{1}{2}\eta_{2K}\eta^K_2)(1-\frac{1}{2}\eta_{3M}\eta^M_3)]  \nonumber \\
& \propto \delta^{(2)}(\mathcal Q^\dag)\sum\limits_{i=1}^3 \lambda^\alpha_{iI}\frac{\partial}{\partial \eta_{jI}}\Big(\lambda_{1,\alpha}(1-\frac{1}{2}\eta_{2K}\eta^K_2)(1-\frac{1}{2}\eta_{3M}\eta^M_3) \nonumber \\
& \quad -\eta_1 \lambda_{2I,\alpha}\eta^I_2(1-\frac{1}{2}\eta_{3M}\eta^M_3) -\eta_1 (1-\frac{1}{2}\eta_{2J}\eta^J_2) \lambda_{3I,\alpha}\eta^I_3 \Big) \nonumber \\
& \propto \delta^{(2)}(\mathcal Q^\dag)( [1 2_I]\eta^I_2 + [1 3_I]\eta^I_3 + m_2\, \eta_1 + m_3 \eta_1) 
\end{align}
where we've pulled out all common factors and abused notation slightly, using $\lambda_{iI}\frac{\partial}{\partial \eta_{jI}}$ to stand for both the massless and massive variants of $\mathcal Q$. If we instead grab the anti-chiral piece of $\Psi_1$, sending $1-\frac{1}{2}\eta_{1K}\eta^K_1 \to 1$:
\begin{align}
\label{eq:m1tozeroachiral}
\mathcal A(\Phi^\dag_1, \Psi_2, \Psi_3) &\propto \delta^{(2)}(\mathcal Q^\dag) (\sum\limits_{i=1}^3 \lambda_{iI}\frac{\partial}{\partial \eta_{jI}})^2[(1-\frac{1}{2}\eta_{2K}\eta^K_2)(1-\frac{1}{2}\eta_{3M}\eta^M_3)]  \nonumber \\
& \propto \delta^{(2)}(\mathcal Q^\dag)  \sum\limits_{i=1}^3 \lambda^\alpha_{iI}\frac{\partial}{\partial \eta_{jI}}\Big(\lambda_{2I,\alpha}\eta^I_2\,(1-\frac{1}{2}\eta_{3M}\eta^M_3) + (1-\frac{1}{2}\eta_{2M}\eta^M_2)\lambda_{3I,\alpha}\eta^I_3\Big) \nonumber \\
& \propto \delta^{(2)}(\mathcal Q^\dag) (m_2 + m_3)
\end{align}
As expected, Eqs.~\eqref{eq:m1tozerochiral} and \eqref{eq:m1tozeroachiral} maintain the little group scaling for each leg.

Taking things further and sending both $m_1, m_2 \to 0$:
\begin{align}
\mathcal A(\Phi_1, \Phi_2, \Psi_3) &\propto \delta^{(2)}(\mathcal Q^\dag) (\sum\limits_{i=1}^3 \lambda_{iI}\frac{\partial}{\partial \eta_{jI}})^2[\eta_1\,\eta_2\,(1-\frac{1}{2}\eta_{3M}\eta^M_3)]  \nonumber \\
& \propto \delta^{(2)}(\mathcal Q^\dag)\sum\limits_{i=1}^3 \lambda^\alpha_{iI}\frac{\partial}{\partial \eta_{jI}} \Big( (\lambda_{1,\alpha}\eta_2 - \eta_1 \lambda_{2,\alpha})(1-\frac{1}{2}\eta_{3M}\eta^M_3) + \eta_1\eta_2 \lambda_{3I,\alpha}\eta^I_3 \Big) \nonumber \\
& \propto \delta^{(2)}(\mathcal Q^\dag)[12] \\
\mathcal A(\Phi_1, \Phi^\dag_2, \Psi_3) &\propto \delta^{(2)}(\mathcal Q^\dag) (\sum\limits_{i=1}^3 \lambda_{iI}\frac{\partial}{\partial \eta_{jI}})^2[\eta_1(1-\frac{1}{2}\eta_{3M}\eta^M_3)]  \nonumber \\
& \propto \delta^{(2)}(\mathcal Q^\dag) \sum\limits_{i=1}^3 \lambda^\alpha_{iI}\frac{\partial}{\partial \eta_{jI}} \Big( \lambda_{1,\alpha}(1-\frac{1}{2}\eta_{3M}\eta^M_3) - \eta_1\lambda_{3I,\alpha}\eta^I_3 \Big) \nonumber \\
& \propto \delta^{(2)}(\mathcal Q^\dag)  ([1 3_I]\eta^I_3 + m_3 \eta_1) \\
\mathcal A(\Phi^\dag_1, \Phi^\dag_2, \Psi_3) &\propto \delta^{(2)}(\mathcal Q^\dag) (\sum\limits_{i=1}^3 \lambda_{iI}\frac{\partial}{\partial \eta_{jI}})^2(1-\frac{1}{2}\eta_{3M}\eta^M_3)  \nonumber \\
& \propto \delta^{(2)}(\mathcal Q^\dag) m_3
\end{align}
While $\mathcal A(\Phi^\dag, \Phi_2, \Psi_3)$ and $\mathcal A(\Phi^\dag_1, \Phi^\dag_2, \Psi_3) $ look almost the same, the $\delta^{(2)}(\mathcal Q^\dag)$ in the two instances will differ. If we further take the third leg to be massless, there are only two nonzero superamplitudes $\mathcal A(\Phi^\dag_1, \Phi^\dag_2,\Phi^\dag_3)$ and $\mathcal A(\Phi_1, \Phi_2, \Phi_3)$ due to the special three-leg helicity and kinematic constraints. However, both of these two come from superpotential contributions and take special forms:
\begin{align}
&\mathcal A(\Phi^\dag_1, \Phi^\dag_2,\Phi^\dag_3)\propto \delta^{(2)}(\mathcal Q^\dag)\\
&\mathcal A(\Phi_1, \Phi_2, \Phi_3)\propto \widetilde{\delta^{(1)}(\mathcal{Q})},
\end{align}
where $\widetilde{\delta^{(1)}(\mathcal{Q})}=[12]\eta_3+[23]\eta_1+[31]\eta_2$ is the first order Grassmann delta function as the Fourier transform of $\delta^{(2)}(\mathcal Q^\dag)$ in the $\eta^\dag$ basis.

\bibliographystyle{utphys}
\bibliography{ref}
\end{document}